\documentclass[prb,showpacs,twocolumn,aps,superscriptaddress,floatfix]{revtex4}
\usepackage{amsmath}
\usepackage{amssymb}
\usepackage{bm}
\usepackage{graphicx}
\usepackage{color}
\usepackage{ulem}

\DeclareMathOperator{\arch}{acosh}
\begin{document}

\title{Antiferromagnetic states and phase separation in doped AA-stacked
graphene bilayers}

\author{A.O. Sboychakov}
\affiliation{CEMS, RIKEN, Saitama, 351-0198, Japan}
\affiliation{Institute for Theoretical and Applied Electrodynamics, Russian
Academy of Sciences, 125412 Moscow, Russia}

\author{A.V. Rozhkov}
\affiliation{CEMS, RIKEN, Saitama, 351-0198, Japan}
\affiliation{Institute for Theoretical and Applied Electrodynamics, Russian Academy of Sciences, 125412 Moscow, Russia}

\author{A.L. Rakhmanov}
\affiliation{CEMS, RIKEN, Saitama, 351-0198, Japan}
\affiliation{Institute for Theoretical and Applied Electrodynamics, Russian Academy of Sciences, 125412 Moscow, Russia}
\affiliation{Moscow Institute of Physics and Technology, Dolgoprudnyi, Moscow Region, 141700 Russia}

\author{Franco Nori}
\affiliation{CEMS, RIKEN, Saitama, 351-0198, Japan}
\affiliation{Department of Physics, University of Michigan, Ann Arbor, MI 48109-1040, USA}

\begin{abstract}
We study electronic properties of AA-stacked graphene bilayers. In the
single-particle approximation such a system has one electron band  and one
hole band crossing the Fermi level. If the bilayer is undoped, the Fermi
surfaces of these bands coincide. Such a band structure is unstable with
respect to a set of spontaneous symmetry violations. Specifically, strong
on-site Coulomb repulsion stabilizes antiferromagnetic order. At small
doping and low temperatures, the homogeneous phase is unstable, and
experiences phase separation into an undoped antiferromagnetic insulator
and a metal. The metallic phase can be either antiferromagnetic
(commensurate or incommensurate) or paramagnetic depending on the system
parameters. We derive the phase diagram of the system on the
doping-temperature plane and find that, under certain conditions, the
transition from paramagnetic to antiferromagnetic phase may demonstrate
re-entrance. When disorder is present, phase separation could manifest
itself as a percolative insulator-metal transition driven by doping.
\end{abstract}


%
%
%
%
%
%
%
%
%
%

\maketitle

\section{Introduction}

Graphene is the first experimentally-realizable stable true atomic
monolayer. It has a host of unusual electronic properties
\cite{castro_neto_review2009,chakraborty_review,meso_review}.
After its discovery, physical properties of graphene became the subject of
intense scientific efforts. In addition to single-layer graphene, bilayer
graphene is also actively studied. This interest is driven by the desire to
extend the family of graphene-like materials and to create materials with a
gap in the electronic spectrum, which could be of interest for
applications.

\begin{figure}
\centering
\includegraphics[width=0.85\columnwidth]{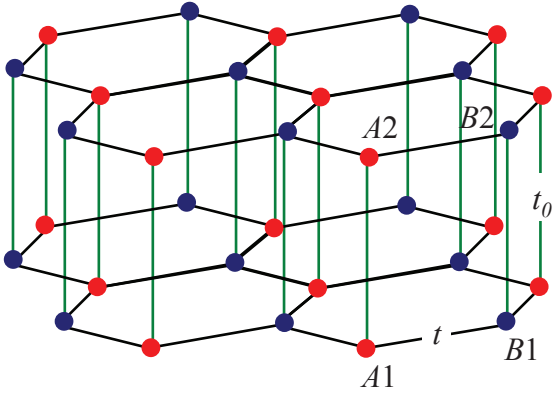}
\caption{(Color online) Crystal structure of the AA-stacked bilayer
graphene. The circles denote carbon atoms in the
${\cal A}$
(red) and
${\cal B}$
(blue) sublattices in the bottom (1) and top (2) layers. The unit cell of
the AA-BLG consists of four atoms $A1$, $A2$, $B1$, and $B2$.  Hopping
integrals $t$ and $t_0$ correspond to the in-plane and inter-plane
nearest-neighbor hopping.
\label{AABLG}}
\end{figure}

Bilayer graphene exists in two stacking modifications. The most common is
the so-called Bernal, or AB, stacking of graphene bilayers (AB-BLG). In
such a stacking, half of the carbon atoms in the top layer are located
above the hexagon centers in the lower layer; and half of the atoms in the
top layer lie above the atoms in the lower layer. A different layer
arrangement, in which carbon atoms in the upper layer are located on top of
the equivalent
atoms of the bottom layer, is referred to as AA-stacked graphene bilayer
(AA-BLG),
Fig.~\ref{AABLG}.
So far, the most efforts have been focused on studying the
AB-BLG~\cite{mccann2006},
for which high-quality samples are
available~\cite{susp_bilayer2009,Mayorov_Sci2011}.
In recent years, the experimental realization of the AA-BLG has been also
reported~\cite{aa first,aa_experiment2008,borysiuk_aa2011}.
However, AA-BLG received limited amount of theoretical attention
\cite{aa_dft2008,spin-orbit2011,borysiuk_aa2011,aa_adsorbtion2010,
aa_optics_2010}.

The tight-binding analysis shows that both AA and AB-BLGs have four bands
(two hole bands and two electron bands). However, the structure of these
bands is different. In the undoped AB-BLG, two bands (one hole band and one
electron band) touch each other at two Fermi points, and the low-energy
band dispersion is nearly
parabolic~\cite{ABBLG}.
The AA-BLG has two bands near the Fermi energy, one electron-like and one
hole-like~\cite{aa_dft2008,spin-orbit2011}.
The low-energy dispersion in the AA-BLG is linear, similar to the monolayer
graphene. Unlike the latter, however, AA-BLG have Fermi surfaces instead of
Fermi points.

An important feature of the AA-BLG is that the hole and electron Fermi
surfaces coincide in the undoped material. It was shown in
Ref.~\onlinecite{our_preprint}
that these degenerate Fermi surfaces are unstable when an arbitrarily weak
electron interaction is present, and the bilayer becomes an
antiferromagnetic (AFM) insulator with a finite electron gap. This
electronic instability is strongest when the bands cross at the Fermi
level. Doping shifts the Fermi level and suppresses the AFM
instability~\cite{our_preprint2}.
Assuming a homogeneous ground state, here we demonstrate that the AFM gap
$\Delta$ decreases when the doping $x$ grows, and vanishes for dopings above
some critical value
$x_c$.
However, the homogeneously-doped state, depending on temperature and
doping, may become unstable with respect to phase separation into an undoped
AFM insulator and a doped
metal~\cite{our_preprint2}.
In the phase-separated state, the concentration of the AFM insulator
decreases when doping increases. Above a certain threshold value of doping
$x^*$,
the insulator-to-metal transition occurs.

In this paper we present a detailed study of the electronic properties of
the AA-BLG. In Sec.~\ref{TBH}, we write down its tight-binding model
Hamiltonian and briefly analyze its properties. In Sec.~\ref{CAFM}, we add
the on-site Coulomb interaction to the Hamiltonian and derive the mean-field
equations for the commensurate AFM gap for finite doping and temperature.
The incommensurate AFM state is analyzed in
Sec.~\ref{ICAFM}.
In
Sec.~\ref{PS},
we demonstrate that the AA-BLG is unstable with respect to phase separation
within some doping and temperature range. The obtained results are
discussed in
Sec.~\ref{Discussion}.

\section{Tight-binding Hamiltonian}\label{TBH}

The crystal structure of the AA-BLG is shown in Fig.~\ref{AABLG}. The
AA-BLG consists of two graphene layers, $1$ and $2$. Each carbon atom of the
upper layer is located above the corresponding atom of the lower layer.
Each layer consists of two triangular sublattices ${\cal A}$ and ${\cal
B}$. The elementary unit cell of the AA-BLG contains four carbon atoms
$A1$, $A2$, $B1$, and $B2$.

We write the single-particle Hamiltonian of the AA-BLG in the form
\begin{eqnarray}\label{H0}
H_0&=&-t\sum_{\langle\mathbf{nm}\rangle i\sigma}\left(d^{\dag}_{\mathbf{n}i{\cal A}\sigma}
d^{\phantom{\dag}}_{\mathbf{m}i{\cal B}\sigma}+H.c.\right)-\\
&&t_0\sum_{\mathbf{n}a\sigma}\left(d^{\dag}_{\mathbf{n}1a\sigma}d^{\phantom{\dag}}_{\mathbf{n}2a\sigma}+H.c.\right)
-\mu\sum_{\mathbf{n}ia\sigma}d^{\dag}_{\mathbf{n}ia\sigma}d^{\phantom{\dag}}_{\mathbf{n}ia\sigma}\,.\nonumber
\end{eqnarray}
Here $d^{\dag}_{\mathbf{n}ia\sigma}$
and
$d^{\phantom{\dag}}_{\mathbf{n}ia\sigma}$
are the creation and annihilation operators of an electron with spin
projection $\sigma$ in the layer
$i=1,\,2$
on the sublattice
$a={\cal A},{\cal B}$
at the position
$\mathbf{n}$,
$\mu$ is the chemical potential, and
$\langle ...\rangle$
denotes nearest-neighbor pair. The amplitude $t$ ($t_0$) in
Eq.~\eqref{H0} describes the in-plane (inter-plane) nearest-neighbor
hopping. For calculations, we will use the values of the hopping integrals
$t\approx2.57$\,eV, $t_0\approx0.36$\,eV
computed by DFT for multilayer AA systems in
Ref.~\onlinecite{Charlier}.

If we perform the unitary transformation
\begin{equation}\label{U1}
h^{\phantom{\dag}}_{\mathbf{n}a\sigma}=\frac{d^{\phantom{\dag}}_{\mathbf{n}1a\sigma}+d^{\phantom{\dag}}_{\mathbf{n}2a\sigma}}{\sqrt{2}}\,,\;\;
g^{\phantom{\dag}}_{\mathbf{n}a\sigma}=\frac{d^{\phantom{\dag}}_{\mathbf{n}1a\sigma}-d^{\phantom{\dag}}_{\mathbf{n}2a\sigma}}{\sqrt{2}}\,,
\end{equation}
then, Eq.~\eqref{H0} can be rewritten as
\begin{eqnarray}\label{H01}
&&\!\!\!H_0\!=\!-t\!\!\!\sum_{\langle\mathbf{nm}\rangle\sigma}\!\!\!\left(h^{\dag}_{\mathbf{n}{\cal A}\sigma}
h^{\phantom{\dag}}_{\mathbf{m}{\cal B}\sigma}+H.c.\right)- (\mu+t_0)\!\!\sum_{\mathbf{n}a\sigma}h^{\dag}_{\mathbf{n}a\sigma}h^{\phantom{\dag}}_{\mathbf{n}a\sigma}\nonumber\\
&&\!\!\!-t\!\!\sum_{\langle\mathbf{nm}\rangle\sigma}\!\!\!\left(g^{\dag}_{\mathbf{n}{\cal A}\sigma}
g^{\phantom{\dag}}_{\mathbf{m}{\cal B}\sigma}+H.c.\right)- (\mu-t_0)\!\!\sum_{\mathbf{n}a\sigma}g^{\dag}_{\mathbf{n}a\sigma}g^{\phantom{\dag}}_{\mathbf{n}a\sigma}\,.
\end{eqnarray}
Therefore, in this representation the Hamiltonian $H_0$ is a sum of two
single-layered graphene
Hamiltonians~\cite{castro_neto_review2009},
with different effective chemical potential
$\mu \pm t_0$.

\begin{figure}
\centering
\includegraphics[width=0.95\columnwidth]{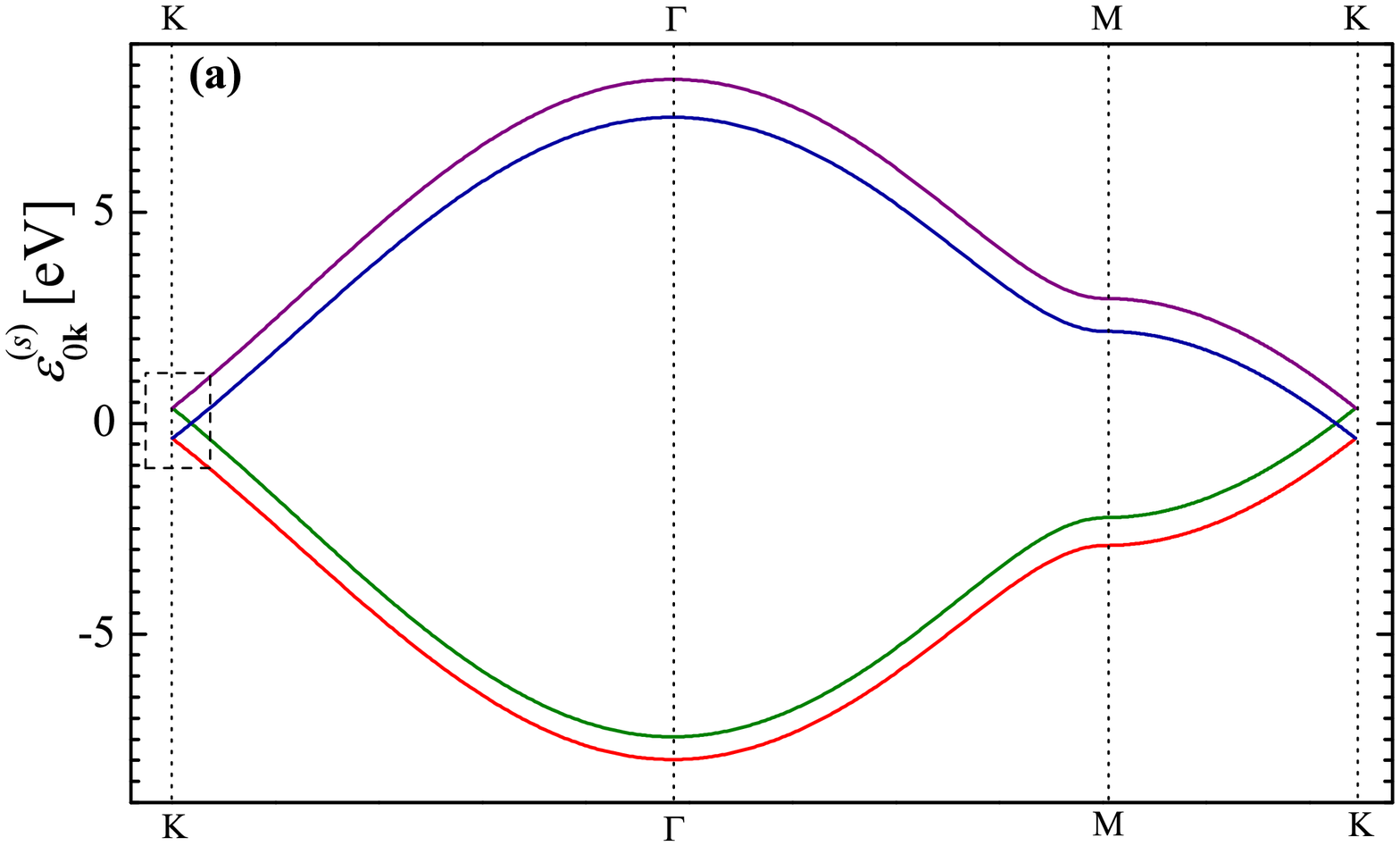}\\
\includegraphics[width=0.95\columnwidth]{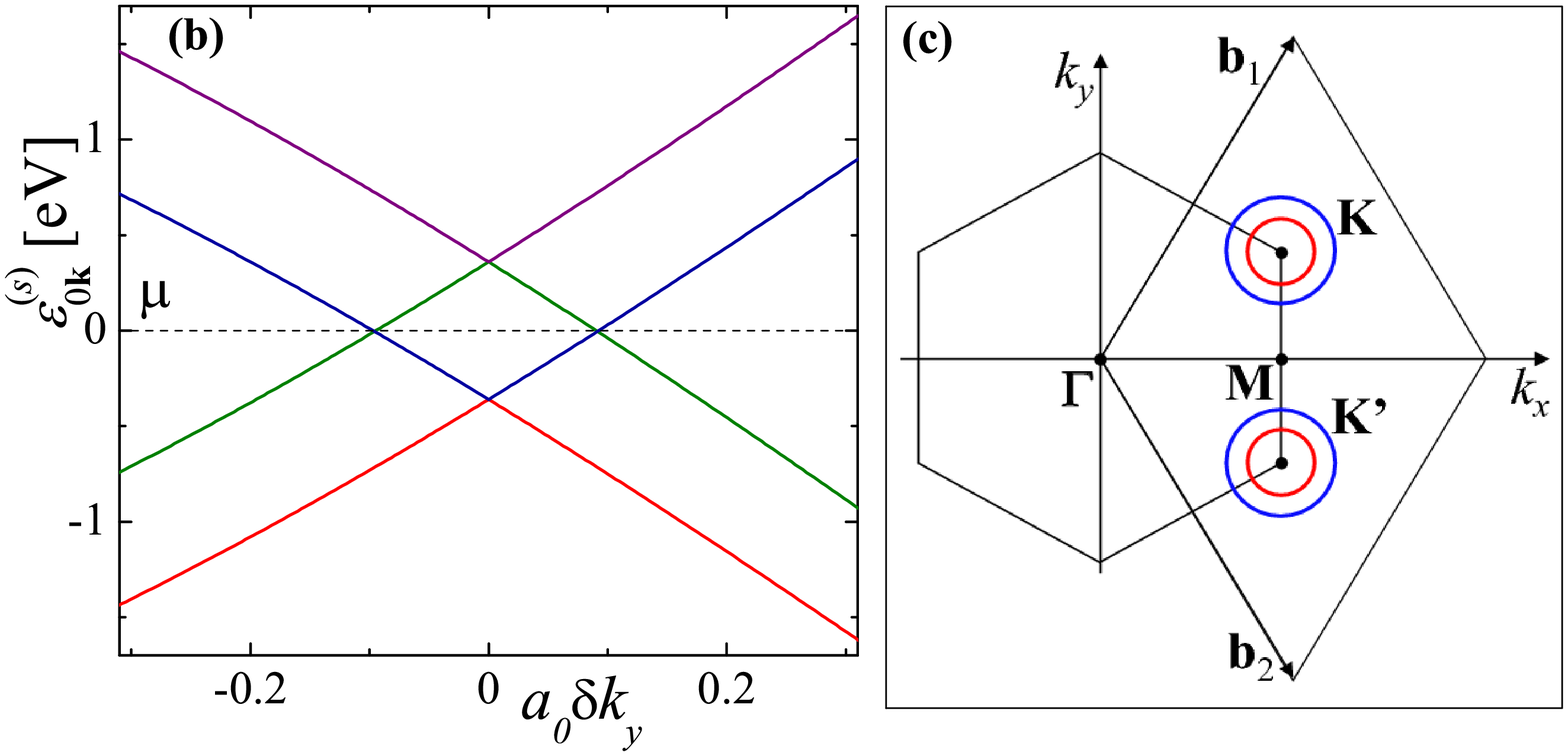}
\caption{(Color online) (a) The single-particle band structure of the
AA-stacked bilayer graphene. It consists of two single-layered graphene
spectra shifted relative to each other by the energy $2t_0$. (b) The
$\mathbf{k}$-dependence of the spectra 
$\varepsilon^{(s)}_{0\mathbf{k}}$ near the Dirac point ${\cal K}$ located
at momentum $\mathbf{K}$.
Here,
$\mathbf{k}=\mathbf{K}+\delta k_y\mathbf{e}_y$.
The intersection of the bands $s=2$ and $s=3$
occurs exactly at zero energy, which corresponds to the Fermi level of the
undoped system. (c) The first Brillouin zone (hexagon) and the reciprocal
lattice unit cell (rhombus) of the AA-BLG. The circles around
${\bf K}$ and ${\bf K}'$ points correspond to Fermi surfaces of the doped system.}
\label{FigSpec0}
\end{figure}

The Hamiltonian~\eqref{H01} can be readily diagonalized. To perform the
diagonalization, we switch
to the fermion operators
$h^{\phantom{\dag}}_{\mathbf{k}a\sigma}$
and
$g^{\phantom{\dag}}_{\mathbf{k}a\sigma}$,
which are defined in the momentum representation, and make the unitary
transformation
\begin{eqnarray}
\gamma^{\phantom{\dag}}_{\mathbf{k}1\sigma}=\frac{h^{\phantom{\dag}}_{\mathbf{k}\cal{A}\sigma}+h^{\phantom{\dag}}_{\mathbf{k}\cal{B}\sigma}e^{i\varphi_{\mathbf{k}}}}{\sqrt{2}}\,,\;
\gamma^{\phantom{\dag}}_{\mathbf{k}2\sigma}=\frac{h^{\phantom{\dag}}_{\mathbf{k}\cal{A}\sigma}-h^{\phantom{\dag}}_{\mathbf{k}\cal{B}\sigma}e^{i\varphi_{\mathbf{k}}}}{\sqrt{2}}\,,\nonumber\\
\gamma^{\phantom{\dag}}_{\mathbf{k}3\sigma}=\frac{g^{\phantom{\dag}}_{\mathbf{k}\cal{A}\sigma}+g^{\phantom{\dag}}_{\mathbf{k}\cal{B}\sigma}e^{i\varphi_{\mathbf{k}}}}{\sqrt{2}}\,,\;
\gamma^{\phantom{\dag}}_{\mathbf{k}4\sigma}=\frac{g^{\phantom{\dag}}_{\mathbf{k}\cal{A}\sigma}-g^{\phantom{\dag}}_{\mathbf{k}\cal{B}\sigma}e^{i\varphi_{\mathbf{k}}}}{\sqrt{2}}\,,
\end{eqnarray}
where
$\varphi_{\mathbf{k}}=\arg \left(f_{\mathbf{k}}\right)$,
\begin{equation}\label{f}
f_{\mathbf{k}}
=
1
+2\exp\!\left(\frac{3ik_xa_0}{2}\right)
\!
\cos\!\left(\!\!\frac{\sqrt{3} k_ya_0}{2}\!\!\right)\,,
\end{equation}
and $a_0$ is the in-plane carbon-carbon distance. As a result,
Hamiltonian~\eqref{H01}
becomes
\begin{equation}
\label{H0diag}
H_0=\!\sum_{\mathbf{k}s\sigma}\!
\left(\varepsilon^{(s)}_{0\mathbf{k}}-\mu\right)
\gamma^{\dag}_{\mathbf{k}s\sigma}
\gamma^{\phantom{\dag}}_{\mathbf{k}s\sigma}\,.
\end{equation}
In this equation, the band index $s$ runs from 1 to 4, and the band spectra
$\varepsilon^{(s)}_{0\mathbf{k}}$
are
\begin{eqnarray}
\label{E0k}
&&\varepsilon^{(1)}_{0\mathbf{k}}=-t_0-t\zeta_{\bf k}\,,\qquad
\varepsilon^{(2)}_{0\mathbf{k}}=-t_0+t\zeta_{\bf k}\,,\nonumber\\
&&\varepsilon^{(3)}_{0\mathbf{k}}=+t_0-t\zeta_{\bf k}\,,\qquad
\varepsilon^{(4)}_{0\mathbf{k}}=+t_0+t\zeta_{\bf k}\,,
\end{eqnarray}
where
$\zeta_{\bf k} = |f_{\bf k}|$.
The band structure obtained is shown in Fig.~\ref{FigSpec0}. The bands
$s=2$ and $s=3$ cross the Fermi level near the Dirac points
$\mathbf{K}=2\pi (\sqrt{3},\,1 )/(3\sqrt{3}a_0)$
and
$\mathbf{K}'=2\pi (\sqrt{3},\,-1 )/(3\sqrt{3}a_0)$
[see
Fig.~\ref{FigSpec0}(b)].

For undoped systems
($\mu = 0$, half-filling)
the Fermi surfaces are given by the equation
$|f_{\bf k}|=t_0/t$.
Since
$t_0/t\ll1$,
we can expand the function
$|f_{\bf k}|$
near the Dirac points and find that the Fermi surface consists of two
circles with radius
$k_{r}=2t_0/(3ta_0)$.
These Fermi surfaces transform into four circles in doped AA-BLG [see
Fig.~\ref{FigSpec0}(c)].

The most important feature of this tight-binding band structure is that at
half-filling the Fermi surfaces of both bands coincide. That is, the
electron and hole components of the Fermi surface are perfectly nested.
This property of the Fermi surfaces is quite stable against changes in the
tight-binding Hamiltonian. It survives even if longer-range hoppings are
taken into account, or a system with two non-equivalent layers is
considered (e.g., similar to the single-side hydrogenated
graphene~\cite{sshg}).
However, the electron interactions can destabilize such a degenerate
spectrum, generating a
gap~\cite{our_preprint}.

\section{Commensurate antiferromagnetic state}\label{CAFM}


The single-electron spectrum described in the previous section
changes qualitatively when interaction is included. Specifically, using
mean field theory, we will demonstrate that the degenerate Fermi surface of
the undoped AA-BLG is unstable with respect to the spontaneous generation
of AFM order.

\subsection{Mean-field equations}\label{CAFMA}

We approximate the electron-electron interaction by the Hubbard-like
interaction Hamiltonian:
\begin{equation}\label{U}
H_{\text{int}}=\frac{U}{2}\sum_{\mathbf{n}ia\sigma}
\left(n_{\mathbf{n}ia\sigma}-\frac{1}{2}\right)\left(n_{\mathbf{n}ia\bar{\sigma}}-\frac{1}{2}\right)\,,
\end{equation}
where
$n_{\mathbf{n}ia\sigma}=d^{\dag}_{\mathbf{n}ia\sigma}d^{\phantom{\dag}}_{\mathbf{n}ia\sigma}$, and $\bar{\sigma}=-\sigma$.
It is known that the on-site Coulomb interaction $U$ in graphene and other
carbon systems is rather strong, but the estimates available in the
literature vary
considerably~\cite{Ut,U69},
ranging from 4-5 to 9-10~eV.

We analyze the properties of the Hamiltonian $H=H_{0}+H_{\text{int}}$ in
the mean-field approximation. We choose the $x$-axis as the spin
quantization axis, and write the order parameters as
\begin{eqnarray}
\Delta_{ia}\equiv U\left\langle
d^{\dag}_{\mathbf{n}ia\uparrow}
d^{\phantom{\dag}}_{\mathbf{n}ia\downarrow}
\right\rangle\,,
\\
\label{GtypeDelta}
\Delta_{1\cal{A}}
=
\Delta_{2\cal{B}}=-\Delta_{1\cal{B}}=-\Delta_{2\cal{A}}\equiv\Delta\,,
\end{eqnarray}
and $\Delta$ is real. Such AFM order, when spin at any given site is
antiparallel to spins at all four nearest-neighbor sites, is called in the
literature G-type AFM. Other types of spin order are either unstable or
metastable.

In the mean-field approximation, the interaction Hamiltonian has the form
\begin{eqnarray}
\label{UMF}
H^{\rm MF}_{\text{int}}
&=&
{\cal N}\left[
		\frac{4\Delta^2}{U}-U(n^2-1)
       \right]
+
\frac{Ux}{2}
\sum_{\mathbf{n}ia\sigma}n_{\mathbf{n}ia\sigma}
\nonumber\\
&-&\sum_{\mathbf{n}ia}
	\Delta_{ia}
	\left(
		d^{\dag}_{\mathbf{n}ia\uparrow}
		d^{\phantom{\dag}}_{\mathbf{n}ia\downarrow}
		+
		d^{\dag}_{\mathbf{n}ia\downarrow}
		d^{\phantom{\dag}}_{\mathbf{n}ia\uparrow}
	\right)
\,,
\end{eqnarray}
where $x=n-1$ is the doping level, $n$ is the number of electrons per site,
and ${\cal N}$ is the number of unit cells in the sample (a unit cell of
AA-BLG consists of four carbon atoms, see
Fig.~\ref{AABLG}). 
Below, when quoting numerical estimates for doping, we will write $x$ as a
percentage of the total number of carbon atoms in the sample.

We introduce the four-component spinor
\begin{eqnarray}
\psi^{\dag}_{\mathbf{k}\sigma}
=
(
	d^{\dag}_{\mathbf{k}1\cal{A}\sigma},
	d^{\dag}_{\mathbf{k}2\cal{A}\sigma},
	d^{\dag}_{\mathbf{k}1\cal{B}\sigma},
	d^{\dag}_{\mathbf{k}2\cal{B}\sigma}
),
\end{eqnarray}
which can be used to build an eight-component spinor
$\Psi^{\dag}_{\mathbf{k}}
=
(
	\psi^{\dag}_{\mathbf{k}\uparrow},
	\psi^{\dag}_{\mathbf{k}\downarrow}
).$
In terms of this spinor, the mean field Hamiltonian
$H^{\rm MF}=H_{0}+H^{\rm MF}_{\text{int}}$
can be written as
\begin{eqnarray}
\label{HtotM}
H^{\rm MF}
=
{\cal N}E_0
+
\sum_{\mathbf{k}}
	\Psi^{\dag}_{\mathbf{k}}
	\left(
		\begin{matrix}
			\hat{H}_{0\mathbf{k}} - \mu' &\hat{\Delta}&\cr
			\hat{\Delta}&\hat{H}_{0\mathbf{k}} - \mu' \cr
		\end{matrix}\!\!\!
	\right)
	\Psi^{\phantom{\dag}}_{\mathbf{k}}\,,
\end{eqnarray}
where
\begin{eqnarray}
E_0 = \frac{4\Delta^2}{U} - U(n^2-1),\ \  \mu'=\mu-\frac{Ux}{2}.
\end{eqnarray}
In these equations,
$E_0$
is a $c$-number, $\mu'$ is the renormalized chemical potential,
$\hat{H}_{0\mathbf{k}}$,
and
$\hat{\Delta}$
are
$4\times4$
matrices
\begin{equation}\label{Hk}
\hat{H}_{0\mathbf{k}}=-\left(
\begin{matrix}
0&t_0&tf_{\bf k}&0\cr
t_0&0&0&tf_{\bf k}\cr
tf_{\bf k}^{*}&0&0&t_0\cr
0&tf_{\bf k}^{*}&t_0&0\cr
\end{matrix}\right)\,,
\end{equation}
\begin{equation}\label{DeltaMatr}
\hat{\Delta}=\left(
\begin{matrix}
-\Delta&0&0&0\cr
0&\Delta&0&0\cr
0&0&\Delta&0\cr
0&0&0&-\Delta\cr
\end{matrix}\right)\,.
\end{equation}

We diagonalize the $8\times8$ matrix in Eq.~\eqref{HtotM} and obtain four
doubly-degenerate bands
\begin{eqnarray}\label{Ek}
\varepsilon^{(1,4)}_{\mathbf{k}}
=
\mp\sqrt{\Delta^2+\left(t\zeta_{\mathbf{k}}+t_0\right)^2}\,,
\nonumber\\
\varepsilon^{(2,3)}_{\mathbf{k}}
=
\mp\sqrt{\Delta^2+\left(t\zeta_{\mathbf{k}}-t_0\right)^2}\,.
\end{eqnarray}
To determine the AFM gap $\Delta$ we should minimize the grand potential
$\Omega$. The grand potential per unit cell is
\begin{equation}\label{Omega}
\Omega=E_0-2T\!\sum_{s=1}^{4}\!\int\!\frac{d\mathbf{k}}{V_{\text{BZ}}}\ln\left[1+e^{(\mu'-\varepsilon^{(s)}_{\mathbf{k}})/T}\right]\,,
\end{equation}
where $V_{\text{BZ}}$ is the volume of the first Brillouin zone.

To evaluate integrals over the Brillouin zone it is convenient to introduce
the density of states
\begin{equation}\label{ro}
\rho_0(\zeta)=\!\int\!\frac{d\mathbf{k}}{V_{\text{BZ}}}\delta(\zeta-\zeta_{\mathbf{k}})\,.
\end{equation}
This function is non-zero only for
$0<\zeta<3$.
It is related to the graphene density of states
$\rho_{\text{gr}}(E)$
as
$\rho_{\text{gr}}(E)=\rho_{0}(|E/t|)/t$
(see
Ref.~\onlinecite{castro_neto_review2009}).

Minimization of $\Omega$ with respect to $\Delta$ gives the equation
\begin{eqnarray}
1&=&\frac{U}{4t}\!\int\limits_{0}^{3}\!\!d\zeta\,\rho_0(\zeta)\!\!\label{DeltaT}
\left[F\left(\sqrt{\delta^2+(\zeta+\zeta_0)^2}\right)\right.+\nonumber\\
&&\left.F\left(\sqrt{\delta^2+(\zeta-\zeta_0)^2}\right)\right]\,,
\end{eqnarray}
where  $\delta=\Delta/t$, $\zeta_0=t_0/t$, and
\begin{equation}
F(\varepsilon)
=
\frac{f(-t\varepsilon - \mu')-f(t\varepsilon - \mu')}{\varepsilon},\;\;
f(E)=\frac{1}{e^{\frac{\scriptstyle E}{\scriptstyle T}}+1}\,.
\end{equation}
Equation~\eqref{DeltaT} determines the gap $\Delta$ as a function of the
renormalized chemical potential $\mu'$. To find $\Delta$ as a function of
doping, we need to relate the doping and the chemical potential. It is easy
to prove that
\begin{eqnarray}
n=1+x=-\frac14\frac{\partial(\Omega - E_0)}{\partial\mu'}.
\end{eqnarray}
Then, using Eqs.~\eqref{Omega} and \eqref{ro} we derive
\begin{eqnarray}
x=\frac12\!\int\limits_{0}^{3}\!\!d\zeta\,\rho_0(\zeta)\!\!\label{xT}
\left[G\left(\sqrt{\delta^2+(\zeta+\zeta_0)^2}\right)\right.+\nonumber\\
\left.G\left(\sqrt{\delta^2+(\zeta-\zeta_0)^2}\right)\right]\,,
\\
{\rm where\ \ }
G(\varepsilon)=f(-t\varepsilon - \mu')+f(t\varepsilon - \mu')-1\,.
\end{eqnarray}
Solving
Eqs.~\eqref{DeltaT}
and~\eqref{xT}
we obtain the AFM gap
$\Delta(x,T)$
and the chemical potential
$\mu(x,T)$.

The solutions of
Eqs.~\eqref{DeltaT}
and~\eqref{xT}
satisfy the following relations:
$\Delta(-x,T)=\Delta(x,T)$
and
$\mu(-x,T)=-\mu(x,T)$.
They are consequences of the particle-hole symmetry of the model
Hamiltonian. The next-nearest-neighbor hopping breaks this symmetry.
However, our analysis shows that corrections, introduced by these terms, do
not exceed
1--2\%
for the range of parameters characteristic of graphene systems. Assuming
particle-hole symmetry, below we only consider electron doping,
$x>0$.

\begin{figure}
\centering
\includegraphics[width=0.95\columnwidth]{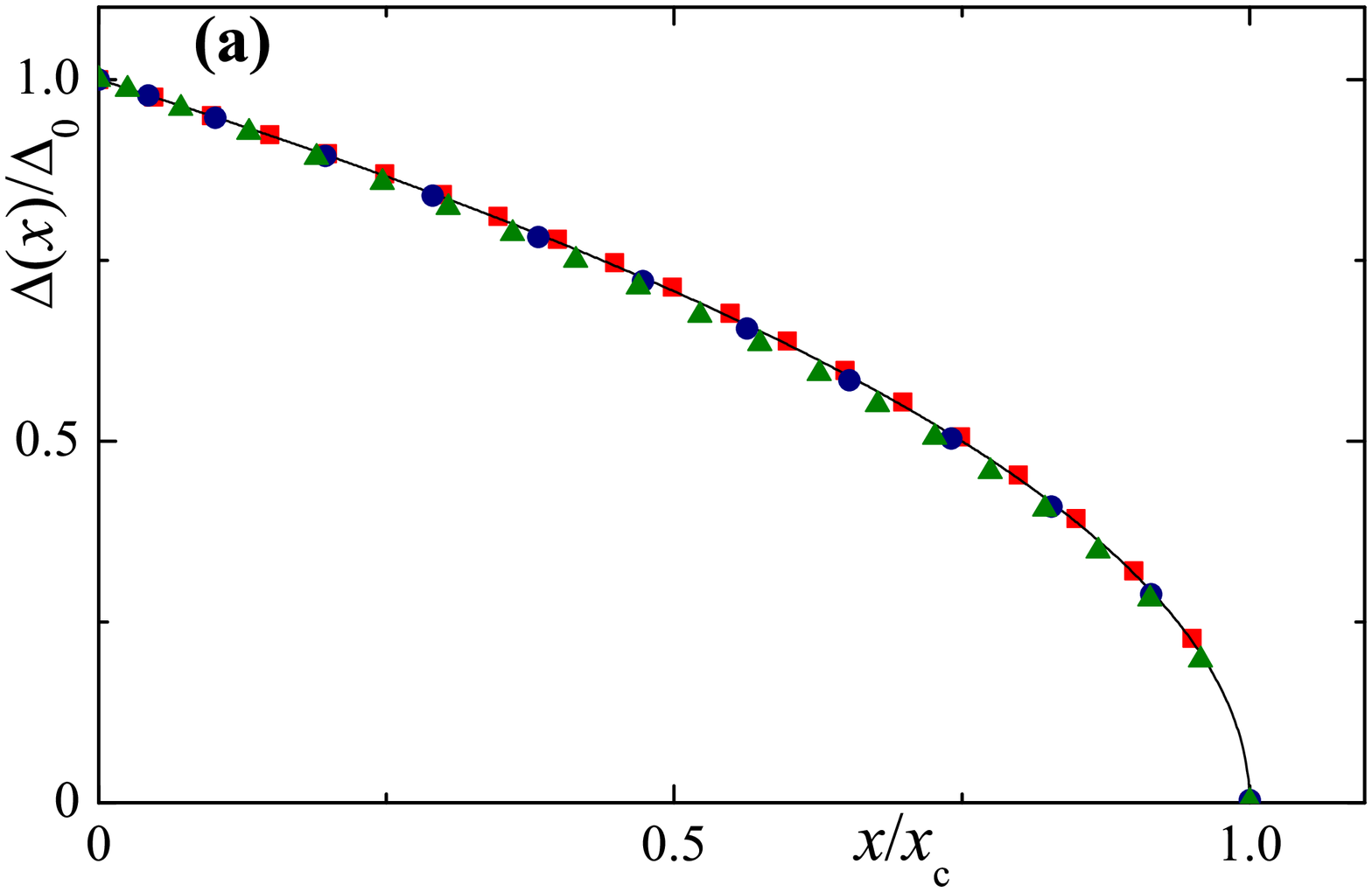}\\
\includegraphics[width=0.95\columnwidth]{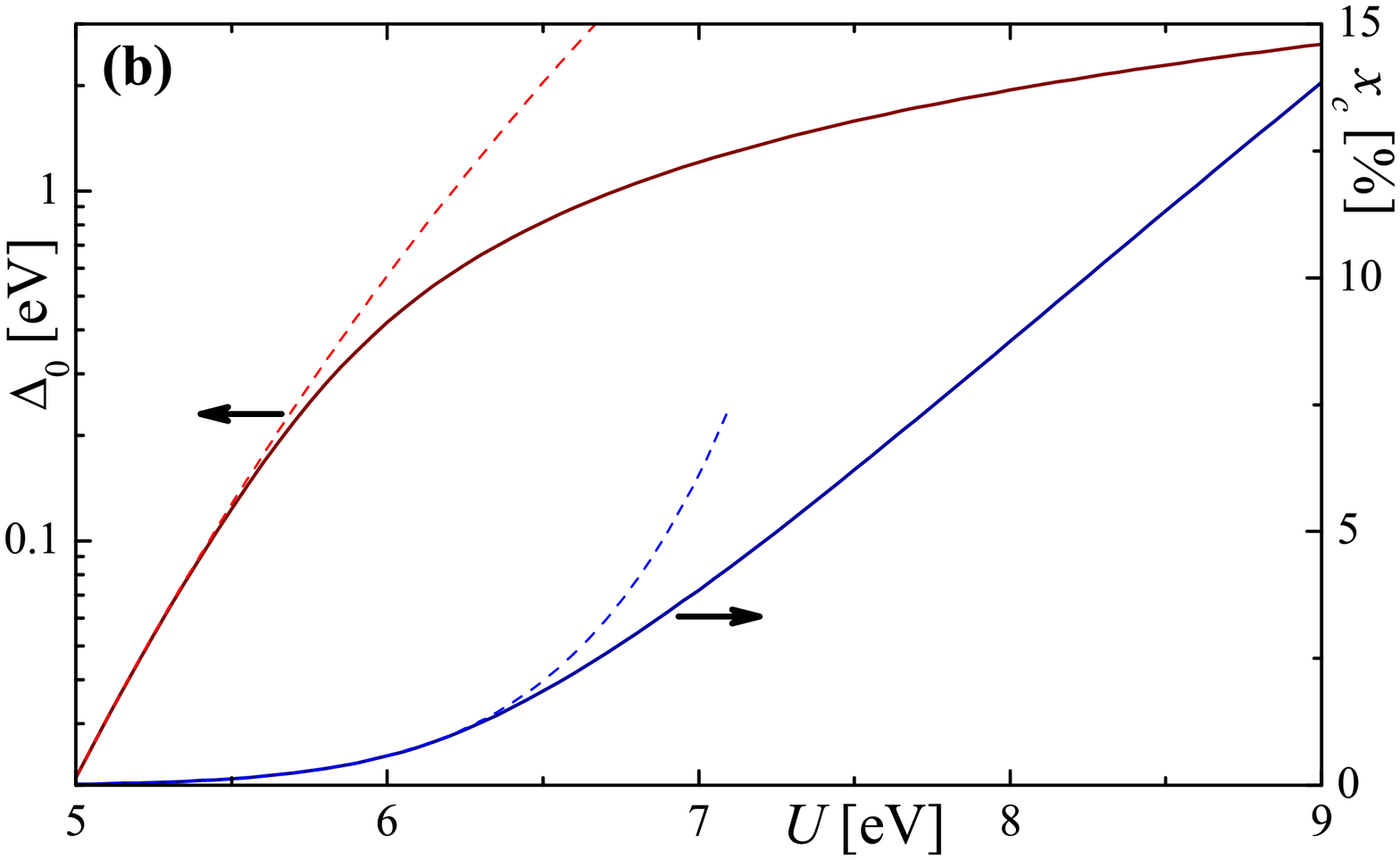}
\caption{(Color online) (a) The AFM gap ratio
$\Delta/\Delta_0$
versus doping
$x/x_c$
for different values of the on-site Coulomb repulsion $U$: (red) squares
correspond to
$U=5.5$\,eV,
(blue) circles to
$U=7$\,eV,
(green) triangles to
$U=9$\,eV.
The solid (black) curve is
$\Delta(x)/\Delta_0=\sqrt{1-x/x_c}$.
(b) The dependencies
$\Delta(x=0,T=0) \equiv \Delta_0$
and critical doping
$x_c$
versus $U$. The solid curves are numerical solutions of
Eqs.~\eqref{EqDelta2}
and
\eqref{EqMu2},
while the dashed curves are calculated using the approximate analytical
solution,
Eqs.~\eqref{xca}
and~\eqref{DeltaAFM}.
}
\label{Gap}
\end{figure}

\subsection{Zero temperature}

If $T=0$, Eqs.~\eqref{DeltaT} and~\eqref{xT} become
\begin{eqnarray}\label{EqDelta2}
1&=&\frac{U}{4t}\int\limits_0^3\!\!d\zeta\,\rho_0(\zeta)\!\!
\left[
	\frac{1-\Theta\left(\displaystyle \mu'/t-
	\sqrt{\delta^2+\left(\zeta+\zeta_0\right)^2}\right)}%
{\sqrt{\delta^2+\left(\zeta+\zeta_0\right)^2}}
\right.+\nonumber\\
&&\left.\frac{1-
\Theta\left(
	\displaystyle \mu'/t-
	\sqrt{\delta^2+\left(\zeta-\zeta_0\right)^2}
      \right)}
{\sqrt{\delta^2+\left(\zeta-\zeta_0\right)^2}}\right]\,,
\end{eqnarray}
\begin{eqnarray}
\label{EqMu2}
x&=&\frac12\int\limits_0^3\!\!d\zeta\,\rho_0(\zeta)\!\!%
\left[\Theta\left(\displaystyle \mu'/t-\sqrt{\delta^2+\left(\zeta+\zeta_0\right)^2}\right)\right.+\nonumber\\
&&\left.
\Theta\left(\displaystyle \mu'/t-\sqrt{\delta^2+\left(\zeta-\zeta_0\right)^2}\right)\right]\,,
\end{eqnarray}
where $\Theta(x)$ is the step function.
At half-filling, $n=1$, $x=0$, and $\mu=\mu'=0$. The lower two bands are filled,
the upper two bands are empty, and both $\Theta$-functions in Eq.~\eqref{EqDelta2} are zero for any $\zeta$.

When doping is introduced, analysis of the latter equations shows that
$\mu'$
changes abruptly from zero to the value
$\mu'>\Delta$.
The gap
$\Delta(x,T=0)$
decreases monotonously from
$\Delta(x=0,T=0) \equiv \Delta_0$
to $0$, when $x$ increases from $0$ to some critical doping $x_c$. To find
$x_c$ we must put
$\delta = 0$
into
Eqs.~\eqref{EqDelta2}
and~\eqref{EqMu2}
and solve them for
$\mu'$
and
$x = x_c$.

Equations~(\ref{EqDelta2})
and
\eqref{EqMu2}
can be solved analytically, if
$\Delta_0\ll t,t_0$.
Using the asymptotic expansions of integrals in these equations for small
$\delta$ we
obtain~\cite{our_preprint2}
\begin{eqnarray}
\label{Delta_vs_x}
\Delta(x,0)&=&\Delta_0\sqrt{1\!-\frac{x}{x_c}}\,,
\\
\label{DelMu}
\mu(x,0)&=&\Delta_0\left[{\rm sgn\,}(x)-\frac{x}{2x_c}\right]+\frac{Ux}{2}\,,
\\
\label{xca}
x_c
&=&
\frac{\Delta_0\rho_0(\zeta_0)}{2t}
\cong
\frac{\Delta_0t_0}{\pi\sqrt{3}t^2}\;\;{\rm \ when\ }t_0\ll t\,.
\end{eqnarray}
In this limit the value of $\Delta_0$ is given by the relation~\cite{our_preprint}
\begin{equation}\label{DeltaAFM}
\Delta_0=2\sqrt{t_0(3t-t_0)}\exp\left\{-\frac{4t-U\eta(\zeta_0)}{2U\rho_0(\zeta_0)}\right\}\,,
\end{equation}
where
\begin{equation}
\eta(\zeta_0)\!=\!\int\limits_0^3\!\!d
\zeta\left[		\frac{\rho_0(\zeta)}{\zeta+\zeta_0}
		+		\frac{\rho_0(\zeta)-\rho_0(\zeta_0)}
		     {\left|\zeta-\zeta_0\right|}\right].\;\;\
\end{equation}

The dependence of the ratio $\Delta(x,0)/\Delta_0$ on $x/x_c$ for different
values of $U$ is shown in Fig.~\ref{Gap}(a). Figure~\ref{Gap}(b) shows
$\Delta_0$ and $x_c$ as functions of $U$ calculated both numerically
[Eqs.~\eqref{EqDelta2} and \eqref{EqMu2}] and analytically
[Eqs.~\eqref{xca} and~\eqref{DeltaAFM}]. Equations~(\ref{DelMu}),
(\ref{xca}) together with Eq.~\eqref{DeltaAFM} for $\Delta_0$ are valid if
$U\lesssim6$\,eV. However,
Eq.~(\ref{Delta_vs_x})
is accurate for any $U$, provided that $x_c$ and $\Delta_0$ are calculated
numerically from
Eqs.~\eqref{EqDelta2}
and~\eqref{EqMu2}
[see
Fig.~\ref{Gap}(a)].

\subsection{Finite temperatures}

In this subsection we will analyze the finite-temperature solutions of the
mean field
equations~(\ref{DeltaT})
and~(\ref{xT}).
However, it is necessary to remember that in 2D systems no long-range order
is possible if
$T>0$.
In such a situation the mean field solutions characterize the short-range
order, which survives for sufficiently low $T$. The effects beyond the mean
field approximation will be discussed in
subsection~\ref{xover}.

\begin{figure}
\centering
\includegraphics[width=0.95\columnwidth]{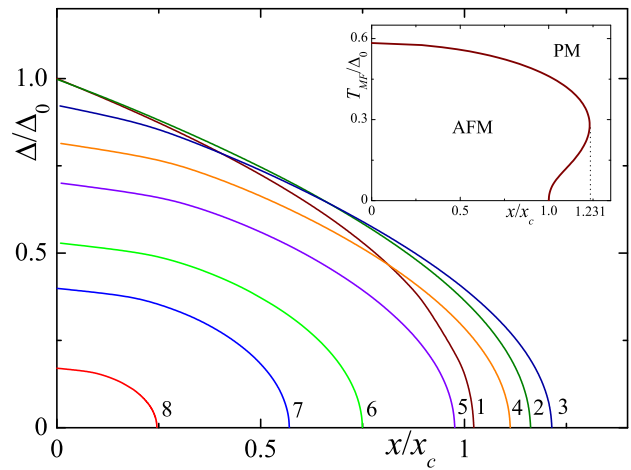}
\caption{(Color online) The dependence of 
$\Delta(x,T)$ 
on doping $x$ calculated for 
$U=5.5$\,eV 
and different 
$T/\Delta_0$: 
(1)~$T/\Delta_0= 0.06$, 
(2)~$T/\Delta_0= 0.17$, 
(3)~$T/\Delta_0= 0.33$, 
(4)~$T/\Delta_0= 0.41$, 
(5)~$T/\Delta_0= 0.47$, 
(6)~$T/\Delta_0= 0.52$, 
(7)~$T/\Delta_0= 0.55$, 
and 
(8)~$T/\Delta_0= 0.58$. 
Inset: The dependence of the mean-field transition temperature on
doping. The re-entrance from PM to AFM state exists in the doping range
$x_c<x<1.231 x_c$; $x_c=0.128$\% and $\Delta_0=0.124$~eV.\label{FigDeltaT}}
\end{figure}

Solving numerically the mean field
equations~(\ref{DeltaT})
and~(\ref{xT}),
we find $\Delta$ as a function of doping $x$ and temperature $T$ (see
Fig.~\ref{FigDeltaT}).
The temperature
$T_{\rm MF}$
at which $\Delta$ vanishes is the mean field transition temperature (see
inset of
Fig.~\ref{FigDeltaT}).
The transition temperature, as a function of doping $x$, is not a
single-valued function. Instead, it demonstrates a pronounced re-entrant
behavior.
We discuss this unusual phenomenon in more details
in Sections \ref{ICAFM}, \ref{PS}, and \ref{Discussion}.

Equations~\eqref{DeltaT}
and
\eqref{xT}
can be simplified in the case of small gap, when
$\Delta_0\ll t_0,t$.
Neglecting terms of the order of
$\Delta_0^2/t^2$
in
Eq.~\eqref{DeltaT}
and taking into account
Eq.~\eqref{DeltaAFM}
for
$\Delta_0$,
we obtain the following equation for $\Delta$
\begin{eqnarray}\label{DeltaTa}
\ln\frac{\Delta_0}{\Delta}
&=&
\frac14\!\!
\int\limits_{\Delta/T}^{\infty}\!\!dz
	\arch\left(\frac{zT}{\Delta}\right)%
\\
\nonumber
&\times&
\!\!\left[
		{\cosh^{-2}\left(\displaystyle\frac{z-\mu'/T}{2}\right)}
		+
		{\cosh^{-2}\left(\displaystyle\frac{z+\mu'/T}{2}\right)}
\right].
\end{eqnarray}
In the same limit, we derive from
Eq.~\eqref{xT}
the relation between $\mu'$ and $x$ in the form
\begin{eqnarray}
\label{xTa}
&&\frac{x}{x_c}
=
\frac{T}{2\Delta_0}\!\!\!
\int\limits_{\Delta/T}^{\infty}\!dz
	\sqrt{z^2-\frac{\Delta^2}{T^2}}\!%
\\
\nonumber
&&\times
	\left[
		{\cosh^{-2}\left(\displaystyle\frac{z-\mu'/T}{2}\right)}
	-
	{\cosh^{-2}\left(\displaystyle\frac{z+\mu'/T}{2}\right)}\right]\,.
\end{eqnarray}
At half-filling ($x=0$) we find from
Eq.~\eqref{DeltaTa}
the BCS-like result
$T_{\rm MF}(0)\cong0.567\Delta_0$.
If we normalize $x$ by
$x_c$
and $\Delta$, $\mu'$ and $T$ by
$\Delta_0$,
then,
Eqs.~\eqref{DeltaTa}
and~\eqref{xTa}
do not include any parameter characterizing the AA-BLG band structure.
Thus, if the electron interaction $U$ is not large, the obtained results do
not depend on details specific to the AA-BLG and are valid for other
systems with imperfect nesting
\cite{RiceMod,our_Rice_model,our_pnic}.

\subsection{Crossover temperature}
\label{xover}

In 2D systems, finite-temperature fluctuations destroy the AFM long-range
order. Then, the results obtained above in the mean-field approximation are
valid only if the mean-field correlation length
$\xi=v_{\rm F}/\Delta$
is smaller than the spin-wave correlation length
$\xi_{sw}$
(here
$v_{\rm F}\cong3a_0t/2$
is the Fermi velocity in our model). Otherwise, short-range ordering of
spins disappears, and we cannot define the AFM order even locally.

In the limit
$\xi_{\rm sw} > \xi$,
the spin fluctuations can be described using the nonlinear $\sigma$-model
with
Lagrangian~\cite{chak,Manousakis}
\begin{eqnarray}
\label{sigma_m}
{\cal L}_{\rm sw}
=\frac{\rho}{2}\left[(\partial_t {\bf D})^2 - c_{\rm sw}^2 (\partial_{\bf r} {\bf D})^2\right],
\end{eqnarray}
where ${\bf D}$ is the unit vector along the local AFM magnetization. The
spin-wave stiffness $\rho$ and velocity
$c_{\rm sw}$
can be evaluated from Eqs.~(7.89) and (7.90) of Ref.~\onlinecite{schakel}
\begin{eqnarray}
c_{\rm sw}=\frac{v_{\rm F}}{\sqrt{2}},\quad\rho=
\begin{cases}
	t_0/(8\pi v_{\rm F}^2), & \text{if $t_0 \gg \Delta$},
	\\
	\Delta/(16\pi v_{\rm F}^2), & \text{if $t_0 \ll \Delta$}.
\end{cases}
\end{eqnarray}
The correlation function
$K(\mathbf{r})=\langle\mathbf{D}(\mathbf{r})\mathbf{D}(0)\rangle$
can be obtained using the
Lagrangian~\eqref{sigma_m}.
At large distances it behaves
as~\cite{chak,Manousakis}
\begin{eqnarray}
K(\mathbf{r})
\approx
1-\frac{T}{\pi\rho v_{\rm F}^2}
\ln\left(\frac{e^{\gamma}\sqrt{2}\,rT}{3a_0t}\right)\,,
\end{eqnarray}
where $\gamma$ is the Euler's constant. The spin-wave correlation length
$\xi_{\rm sw}$
describing the characteristic size of the short-range AFM
order can be estimated using the equation
$K(\xi_{\rm sw})=0$.
Thus, we have
\begin{equation}
\xi_{\rm sw}
\approx
\frac{a_0t}{T}\exp\left(\frac{2\pi\rho c_{\rm sw}^2}{T}\right)\,.
\end{equation}
Solving the equation
$\xi_{\rm sw}=\xi$,
we find the crossover temperature
$T^*$
between the short-range AFM and the PM. The short-range AFM order exists
over distances of about
$\xi_{\rm sw}\gg a_0$,
if
$T<T^*$,
and it is destroyed if
$T>T^*$.
Our numerical analysis shows that
$T^*(x)/T_{\rm MF}(x)\approx0.8$\,-\,$0.9$
for any ratio
$\Delta/t$.
Thus, the mean-field transition temperature gives an appropriate estimate
for the AFM-to-PM crossover temperature.

\section{Incommensurate antiferromagnetic state}\label{ICAFM}

The G-type AFM state considered above has the smallest value of the grand
thermodynamic potential $\Omega$ among other states with commensurate
magnetic order. However, further optimization of $\Omega$ could be achieved
if we allow the local direction of the AFM magnetization slightly rotate
from site to
site~\cite{RiceMod}.
Then, the translation invariance with a lattice period disappears. Such
a state is referred to as incommensurate (or helical) AFM. The complex
order parameter for this state has a form
\begin{equation}\label{Dq}
\Delta_{\mathbf{n}ia}= U\left\langle d^{\dag}_{\mathbf{n}ia\uparrow}d^{\phantom{\dag}}_{\mathbf{n}ia\downarrow}\right\rangle
=e^{i\mathbf{qn}}\Delta_{ia}\,,
\end{equation}
where
$\mathbf{q}$ describes the spatial variation of the AFM magnetization
direction, the position vector
$\mathbf{n}$
specifies the location of a given carbon atom, and
$\Delta_{ia}$
satisfies
Eqs.~\eqref{GtypeDelta}.
The averaged electron spin
$\mathbf{S}_{\mathbf{n}ia}$
at site
$\mathbf{n}$
lies in the $xy$-plane. It is related to the order parameter as
$\langle
d^{\dag}_{\mathbf{n}ia\uparrow}
d^{\phantom{\dag}}_{\mathbf{n}ia\downarrow}
\rangle
=
S^{x}_{\mathbf{n}ia}+iS^{y}_{\mathbf{n}ia}$.
As a result, we obtain
\begin{equation}
\mathbf{S}_{\mathbf{n}ia}
=
\frac{\Delta_{ia}}{U}
\left(
	\cos(\mathbf{qn}),\,\sin(\mathbf{qn})
\right)\,.
\end{equation}

\begin{figure}
\centering
\includegraphics[width=0.95\columnwidth]{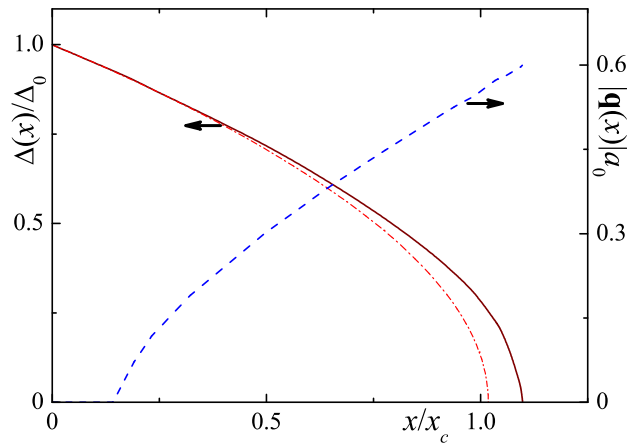}
\caption{(Color online) The dependence of $\Delta$ (red solid curve) and
$|\mathbf{q}|$
(blue dashed curve) on doping $x$ calculated for
$T/\Delta_0=0.06$.
The dot-dashed curve is the gap $\Delta$ calculated for
$\mathbf{q}=0$
and
$U=8$\,eV.
The doping $x$ is normalized by the critical doping
$x_c$,
calculated for the commensurate AFM state. The incommensurate AFM exists in
a slightly larger doping range than the commensurate AFM. Note that the gap
for commensurate AFM remains non-zero even for
$x > x_c$.
This is a manifestation of the re-entrance, see inset of
Fig.~\ref{FigDeltaT}.
\label{FigDeltaQ}}
\end{figure}

The mean-field version of the interaction
Hamiltonian~\eqref{U}
corresponding to the order parameter
$\Delta_{\mathbf{n}ia}$,
Eq.~\eqref{Dq},
can be written in the momentum representation as [c.f.
Eq.~\eqref{UMF}]
\begin{eqnarray}
\label{UMFq}
H_{\text{int}}&\!\!=\!\!&{\cal N}\!\left[\frac{4\Delta^2}{U}-U(n^2-1)\right]+\frac{Ux}{2}\!\sum_{\mathbf{k}ia\sigma}n_{\mathbf{k}ia\sigma}-\\
&&\!\!\sum_{\mathbf{k}ia}\Delta_{ia}\left(d^{\dag}_{\mathbf{k}+\frac{\scriptstyle\mathbf{q}}{\scriptstyle2}ia\uparrow}
d^{\phantom{\dag}}_{\mathbf{k}-\frac{\scriptstyle\mathbf{q}}{\scriptstyle2}ia\downarrow}+
d^{\dag}_{\mathbf{k}-\frac{\scriptstyle\mathbf{q}}{\scriptstyle2}ia\downarrow}
d^{\phantom{\dag}}_{\mathbf{k}+\frac{\scriptstyle\mathbf{q}}{\scriptstyle2}ia\uparrow}\right)\,.\nonumber
\end{eqnarray}
It is convenient to redefine the spinor
$\Psi$
(see Sec.~\ref{CAFMA}):
\begin{eqnarray}
\Psi^{\dag}_{\mathbf{kq}}
=
(
	\psi^{\dag}_{\mathbf{k}+\mathbf{q}/2\uparrow},
	\psi^{\dag}_{\mathbf{k}-\mathbf{q}/2\downarrow}
).
\end{eqnarray}
We can rewrite the mean-field Hamiltonian
$H_0+H_{\rm int}^{\rm MF}$
in the form
\begin{equation}\label{HtotMq}
H
=
{\cal N}E_0
+\!\sum_{\mathbf{k}}
\Psi^{\dag}_{\mathbf{kq}}\!
	\left( \begin{matrix}
		\hat{H}_{
				0\mathbf{k}
				+
				\frac{
					\scriptstyle\mathbf{q}
				     }
				     {
					\scriptstyle2
				     }
			} - \mu'\!\!&\hat{\Delta}&\cr
		\hat{\Delta}&\!\!
		\hat{H}_{
				0\mathbf{k}
				-
				\frac{
					\scriptstyle\mathbf{q}
				     }
				     {
					\scriptstyle2
				     }
			} - \mu'\cr
\end{matrix}\!\!\!\right)\!\Psi^{\phantom{\dag}}_{\mathbf{kq}},
\end{equation}
where
$\hat{H}_{0\mathbf{k}}$
and
$\hat{\Delta}$
are given by
Eqs.~\eqref{Hk}
and
\eqref{DeltaMatr},
respectively.

\begin{figure}
\centering
\includegraphics[width=0.95\columnwidth]{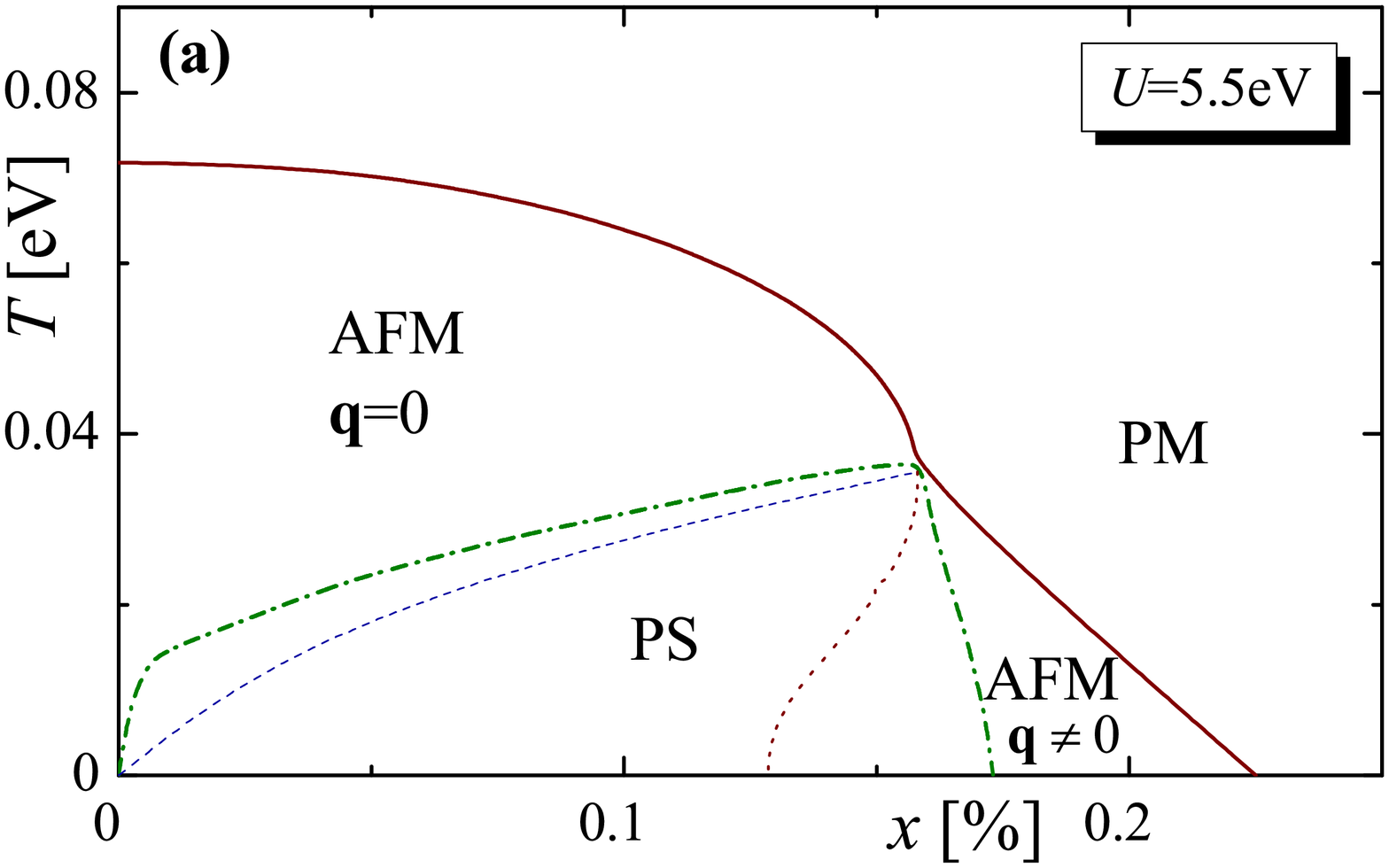}\vspace{0.3cm}\\
\includegraphics[width=0.95\columnwidth]{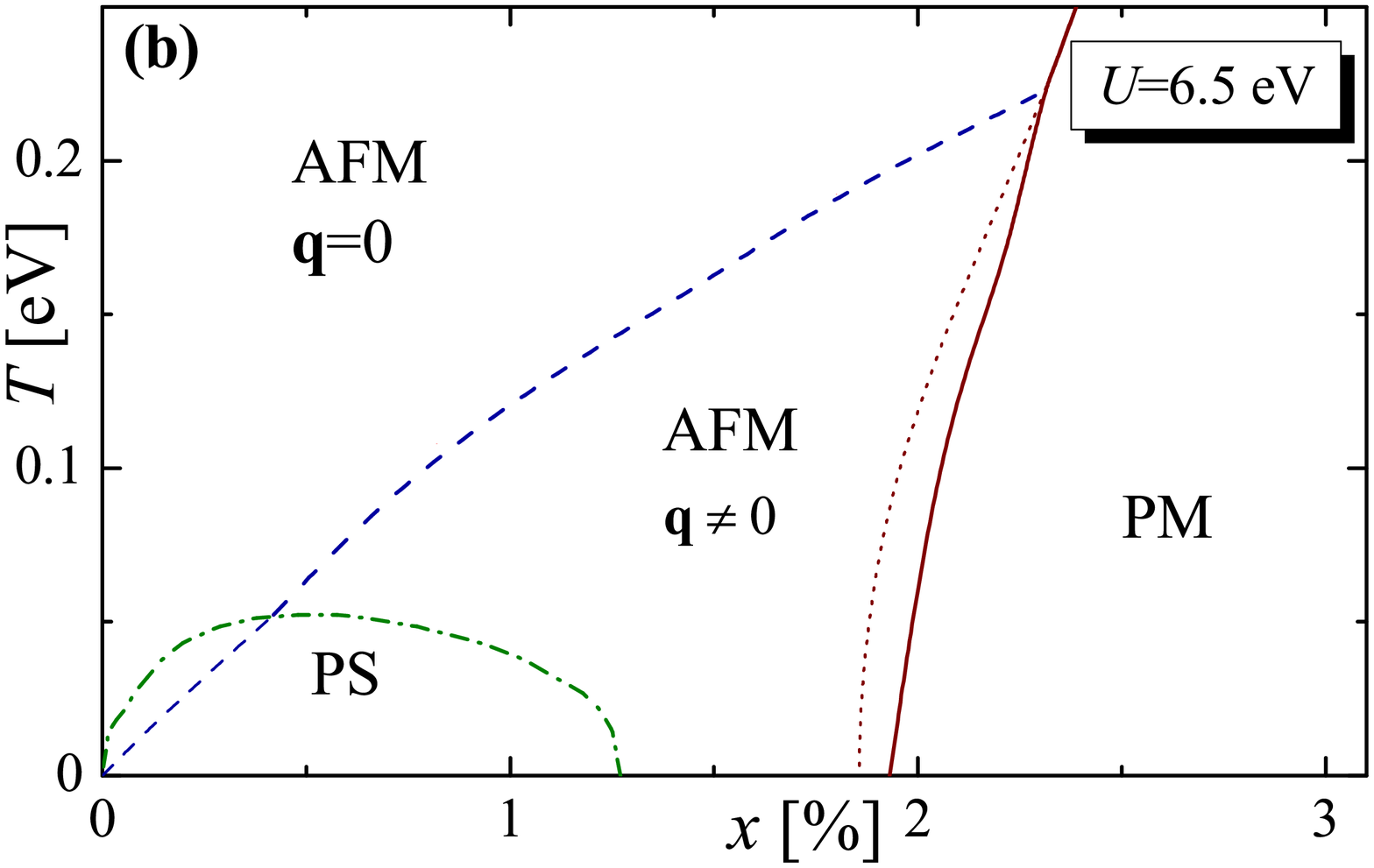}\vspace{0.3cm}\\
\includegraphics[width=0.95\columnwidth]{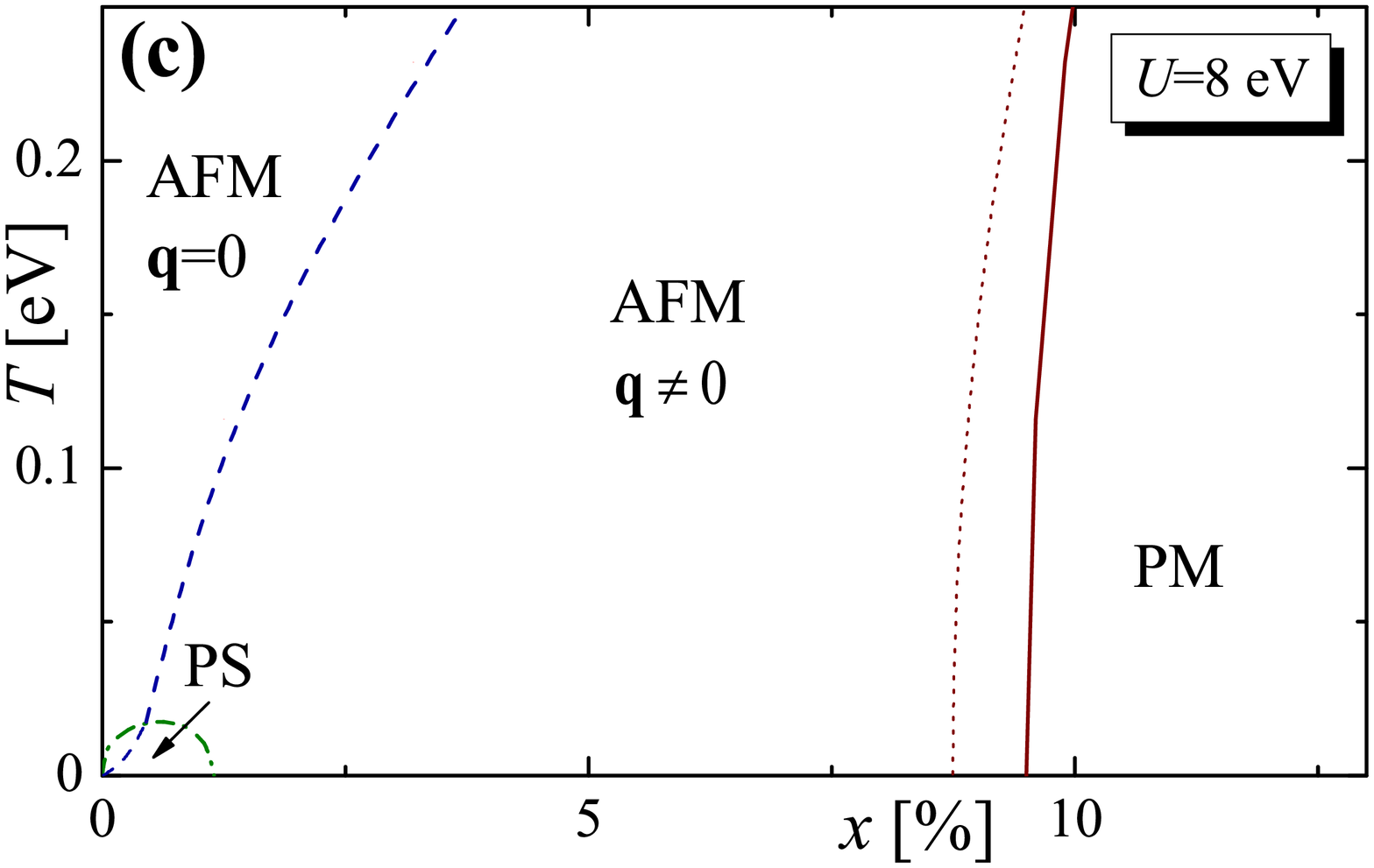}
\caption{(Color online) The phase diagram of the model in the ($x,\,T$)
plane, calculated for the electron doping $x>0$ and
$U=5.5$\,eV (a),
$U=6.5$\,eV (b),
and
$U=8$\,eV (c).
Solid (red) curves are
$T_{\rm MF}(x)$,
(blue) dashed curves are
$T^{q}(x)$,
at which the commensurate-incommensurate transition occurs.
The dotted (red) curves are
$T_{\rm MF}(x)$,
calculated without taking into account the incommensurate AFM state. The
dot-dashed (green) curves show the region of phase separation. For hole
doping
($x<0$)
the results are the same.
\label{FigPhDiagT}}
\end{figure}

The electron spectrum in the incommensurate AFM state is found by
diagonalization of the $8\times8$ matrix in Eq.~\eqref{HtotMq}. It consists
of $8$ non-degenerate bands, $E_{\mathbf{k},\mathbf{q}}^{(s)}$,
$s=1,2,\dots,8$. The analytical expression for
$E_{\mathbf{k},\mathbf{q}}^{(s)}$ can be obtained in the limit
$|\mathbf{q}|\ll1/a_0$,
\begin{equation}
\frac{E_{\mathbf{k},\mathbf{q}}^{(s)}}{t}\!\approx\!\pm\frac{\zeta_{\mathbf{k}+\frac{\scriptstyle\mathbf{q}}{\scriptstyle2}}-
\zeta_{\mathbf{k}-\frac{\scriptstyle\mathbf{q}}{\scriptstyle2}}}{2}\!\pm\!
\sqrt{\frac{\!\Delta^2}{t^2}+\!\!\left[\frac{t_0}{t}\pm\frac{\zeta_{\mathbf{k}+\frac{\scriptstyle\mathbf{q}}{\scriptstyle2}}+
\zeta_{\mathbf{k}-\frac{\scriptstyle\mathbf{q}}{\scriptstyle2}}}{2}\right]^2}\!\!.
\end{equation}
If
$\mathbf{q}=0$,
this spectrum coincides with the spectrum of Eqs.~\eqref{Ek}.

The expression for the grand potential $\Omega$ has similar structure as
Eq.~\eqref{Omega},
but now the summation includes eight bands:
\begin{equation}\label{OmegaQ}
\Omega=E_0-T\sum_{s=1}^{8}\!\int\!\frac{d\mathbf{k}}{V_{\text{BZ}}}\ln\left[1+e^{(\mu'-E^{(s)}_{\mathbf{k},\mathbf{q}})/T}\right].
\end{equation}
Minimization of $\Omega$ with respect to $\Delta$ and
$\mathbf{q}$,
together with the condition relating $x$ and $\mu'$ gives the closed system
of equations for calculating
$\Delta(x,T)$,
$\mathbf{q}(x,T)$,
and
$\mu(x,T)$:
\begin{equation}
\label{SystemQ}
\frac{\partial\Omega}{\partial\Delta}=0\,,\;\;
\frac{\partial\Omega}{\partial\mathbf{q}}=0\,,\;\;
1+x=-\frac{\partial(\Omega - E_0)}{\partial\mu'}\,.
\end{equation}

We calculate the functions $\Delta(x,T)$, $\mathbf{q}(x,T)$, and $\mu(x,T)$
numerically for different values of $U$. Typical curves $\Delta(x)$ and
$|\mathbf{q}(x)|$ are shown in Fig.~\ref{FigDeltaQ}. For comparison, the
curve $\Delta(x)$ calculated for the commensurate AFM is also plotted. We
see that the incommensurate AFM state exists in a slightly wider doping
range than the commensurate one. The incommensurate phase arises at
arbitrary small doping if
$T=0$.
At non-zero $T$ the commensurate AFM state is stable until doping exceeds
some $T$-dependent threshold
$x^{q}(T)$.
The curve $T^{q}(x)$ separates the incommensurate and commensurate AFM
states; the more symmetrical AFM state with
$\mathbf{q}=0$
lies above
$T^{q}(x)$.

The phase diagrams of the model in the $x$--$T$ plane are shown in
Fig.~\ref{FigPhDiagT}
for three different values of $U$. The diagrams for small
$U$ ($\lesssim6$\,eV)
and large
$U$ ($\gtrsim6$\,eV)
demonstrate a qualitative difference. Namely, for small $U$
[Fig.~\ref{FigPhDiagT}(a), $U=5.5$\,eV]
the re-entrance, seen in the inset of
Fig.~\ref{FigDeltaT},
disappears. It is masked by the incommensurate AFM phase. For larger $U$,
however, it survives,
Fig.~\ref{FigPhDiagT}(b,c).
Re-entrance is an unusual phenomenon because the ordering occurs as the
temperature increases. If re-entrance is a genuine feature of the model, or
it is an artifact of the mean field approximation, whose reliability
deteriorates when $U$ grows, we do not know. Similar behavior was predicted
theoretically for quarter-filled Hubbard model at moderate interaction
strength~\cite{theory_reentrance},
and numerically for classical rotor
model~\cite{rotor_reentrance1999}.

\section{Phase separation}\label{PS}

In our discussion above we implicitly assumed that the ground state of the
AA-BLG is spatially homogeneous. However, this is not always true: it was
predicted in
Ref.~\onlinecite{our_preprint2}
that there is a finite doping range where the AA-BLG separates in two
phases with unequal electron densities
$n_{1,2}=1+x_{1,2}$.
Indeed, if
$\Delta_0\ll t,t_0$
we can use
Eq.~\eqref{DelMu}
and obtain that
\begin{eqnarray}
\frac{\partial\mu}{\partial x}<0, {\rm \ \  if\ \ }
\frac{U}{t} < \frac{\pi\sqrt{3}t}{t_0}.
\end{eqnarray}
The negative value of the derivative
$\partial\mu/\partial x$
indicates the instability of the homogeneous state toward phase
separation~\cite{thermodyn}.

If the possibility of the incommensurate AFM is ignored, then a
zero-temperature phase
separation~\cite{our_preprint2}
occurs between the AFM insulator
($x_1=0$)
and the PM
($U\lesssim6$\,eV)
or the AFM
($U\gtrsim6$\,eV)
metal
($x_2>0$).
Here we study phase separation taking into account the incommensurate
AFM phase and non-zero temperature.  We numerically analyze the stability
of the homogeneous state using the dependence of the chemical potential
$\mu$ on the doping $x$.

\begin{figure}
\centering
\includegraphics[width=0.95\columnwidth]{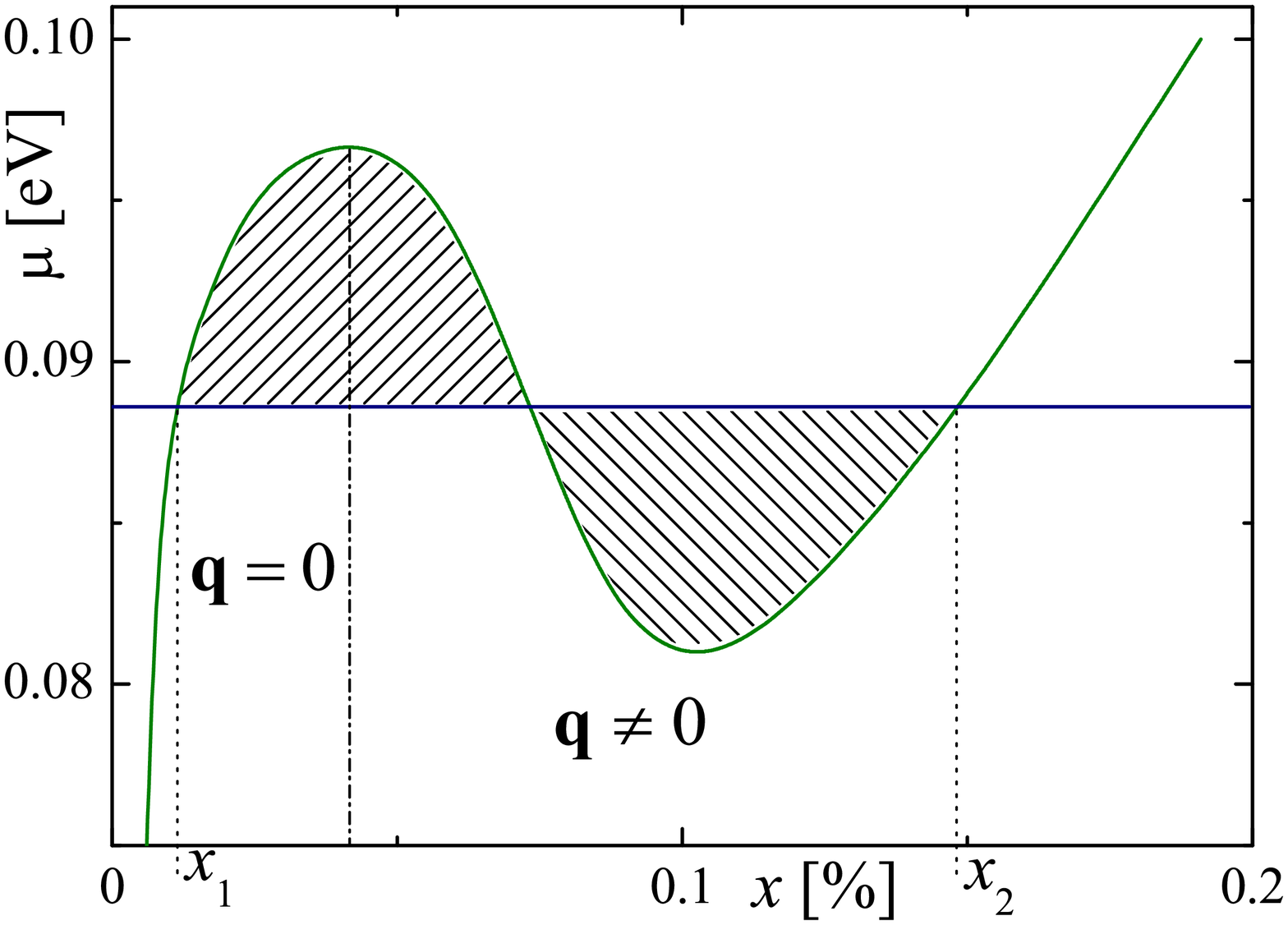}
\caption{(Color online) Chemical potential $\mu$ of the homogeneous state
versus doping $x$;
$U=5.5$\,eV
and
$T=0.014$\,eV.
The vertical dot-dashed line separates the AFM states with
$\mathbf{q}=0$
and
$\mathbf{q}\neq0$.
In the doping range
$x_1<x<x_2$,
phase separation occurs. The values 
$x_{1,2}$ 
are determined by the Maxwell construction: the horizontal (black) line is
drawn in such a manner that the areas of the shaded regions are equal to
each other.
\label{GapMu}
}
\end{figure}

A typical dependence $\mu(x)$ for non-zero temperature is shown in
Fig.~\ref{GapMu}.
The derivative
$\partial\mu/\partial x$
is negative in some range of doping and the system separates in
commensurate
($\mathbf{q}=0$, $x_1<x$)
and incommensurate
($\mathbf{q}\neq0$, $x_2>x$)
AFM phases. The doping concentrations
$x_1$
and
$x_2$
are found using the Maxwell
construction~\cite{thermodyn}:
the (black) horizontal line is drawn in such a manner that the areas of the
shaded regions in
Fig.~\ref{GapMu}
are equal to each other. When temperature increases, the doping range
$x_1<x<x_2$,
where the phase separation exists, becomes narrower and disappears at some critical temperature.
Our calculations show that the separated phases are AFM with
$\mathbf{q}=0$
and
$\mathbf{q}\neq0$
for any values of the model parameters. The region of the phase separation
in the
$(x, T)$-phase diagram is shown in Fig.~\ref{FigPhDiagT} by (green)
dot-dashed lines.

\section{Discussion}
\label{Discussion}

In this paper we study the evolution of the electron properties of AA-BLG
with doping $x$ and temperature $T$.  We calculate the phase diagram of the
system in the
$(x,T)$-plane.
This diagram includes regions of the AFM commensurate and incommensurate
states, a region of phase separation, and the PM state. With good accuracy,
the electronic properties of the AA-BLG are symmetric with respect to the
electron
($x>0$)
or hole
($x<0$)
doping. The maximum crossover temperature between the short-range AFM and
the PM states depends on the on-site Coulomb repulsion $U$.  For example,
AFM ordering can exist up to temperatures of about 50~K if
$U=5.5$\,eV
and to temperatures much higher than room temperature if
$U\gtrsim 6.5$\,eV.
At present, there is no consensus on the value of the on-site Coulomb
repulsion in graphene-based materials.  However, it is commonly accepted
that $U$ lies is in the range
$6<U<10$\,eV
and, consequently, the AFM state can be observed in the AA-BLG.

The critical doping value $x_c$, at which the AFM is replaced by the PM,
also strongly depends on $U$, changing from
$\sim0.1$\,\%
if
$U=5.5$\,eV
to
$\sim10$\,\%
if
$U=8$\,eV.
For graphene systems, doping of about
$10\,\%$
and even higher was
achieved~\cite{CaK}.
Similar to single-layered graphene, the AA-BLG can be doped by using
appropriate
dopants~\cite{CaK,absor},
choosing the substrate and applying a gate
voltage~\cite{Geim,Kim},
or by combinations of these methods.

We predict the existence of phase separation in the AA-BLG. The
separated phases have different electron concentrations, $x_1$ and $x_2$,
and the phase separation will be frustrated by long-range Coulomb
repulsion~\cite{Cul}.  In this case the formation of nano-scale
inhomogeneities is more probable. The electron-rich phase (incommensurate
AFM) is metal and the electron-poor phase (commensurate AFM insulator if
$T=0$) is insulator or ``bad'' metal.  Thus, the percolative
insulator-metal transition will occur when the doping $x$ exceeds some
threshold
value~\cite{our_preprint2},
which is about
$0.5(x_1+x_2)$
in 2D systems. Phase separation exists in the doping range
$x_1<x<x_2$,
and
$x_2\lesssim1$\,\%
for any value of $U$. Depending on $U$, phase separation could be observed
from 30-40~K to room and even higher temperatures (see
Fig.~\ref{FigPhDiagT}).
This makes AA-BLG promising for applications.

The incommensurate AFM phase is mathematically equivalent to the
Fulde-Ferrel-Larkin-Ovchinnikov state in
superconductors~\cite{fflo,sheehy2007},
which is sensitive to
disorder~\cite{Takada}
and difficult to observe experimentally. Consequently, it is reasonable to
expect that the incommensurate AFM phase can be destroyed by factors our
study did not account for. Our calculations predict that in this case the
region of phase separation changes only slightly in the phase diagram.
However, the separated phases would be the AFM insulator and PM
($U\lesssim6$\,eV)
of AFM metal
($U\gtrsim6$\,eV).

To conclude, we studied the phase diagram of the AA-stacked graphene
bilayers on the doping--temperature plane. It consists of paramagnetic and
antiferromagnetic (both commensurate and incommensurate) homogeneous
phases. In addition, a region of phase separation is also identified.
Magnetic properties of the AA-BLG may survive even at room temperature.

\section*{Acknowledgments}

The work was supported by ARO,
Grant-in-Aid for Scientific Research (S),
MEXT Kakenhi on Quantum Cybernetics, and the JSPS via its FIRST program,
the Russian Foundation for Basic Research (projects 11-02-00708,
11-02-00741, 12-02-92100-JSPS, and
12-02-00339). A.O.S. acknowledges support from the RFBR project 12-02-31400
and the Dynasty Foundation.


%

\end{document}